\documentclass[12pt]{article}
\usepackage{graphics}
\usepackage{epsfig}
\usepackage{amsmath}
\title{Multiplicity Fluctuation at Second-order Phase Transition on the Base of Squeezed States}
\author{Babichev L.F., Kuvshinov V.I., Shaparau V.A.\\
\small\textit{Institute of Physics, National Academy
of Sciences of Belarus}\\
\small\textit{F. Skaryna av. 68, 220072 Minsk, Belarus}\\
\small \textit{E-mail: kuvshino@dragon.bas-net.by; shaporov@dragon.bas-net.by}\\
Bukach A.A.\\
\small\textit{Belorussian State University}\\
\small \textit{F. Skaryna av. 4, 220050 Minsk, Belarus}
}

\date{}

\begin{document}

\maketitle
\thispagestyle{empty}
\begin{abstract}
We use the generalized squeezed state description in the framework of the
Ginzburg-Landau theory. Multiplicity distributions of the squeezed
states are studied at second-order phase transition at different
squeeze factors. It is shown that the normalized factorial moments
exhibit a specific behaviour as functions of the resolution scale.
We obtain the values of the scaling exponent which coincides with
experimental data at small squeeze factor.

\vspace{0.2cm}

\vspace{0.2cm}
\end{abstract}
\section{Introduction}
With increasing of the collision energy in $ e^+e^-, p\bar
p, ep,$ heavy-ion experiments the role of the multiparticle production and the collective effects of particle interaction becomes more significant. Large progress has
been achieved in study of the considered processes within
perturbative Quantum Chromodynamics (QCD)
\cite{PQCD}-\cite{Dremin}. Perturbative theory is not able to
reproduce all consequences inherent to corresponding Lagrangian
of the interaction, in particular, the collective aspects of the
behaviour of considered systems as a whole can have
fundamental significance for description of the features of the confinement and hadronization. Here methods of the
statistical physics are fruitful since a mathematical
apparatus used at investigation of the multiplicity distribution
and particle correlations is common both for the statistical
physics and for the processes of the multiparticle production \cite{stat}.
Investigations of the collective excited modes of the hadron
(quark-gluon) medium began to carry out not long ago and have a
phenomenological character.

For statistical systems the fluctuations are large near critical
points. Therefore the multiplicity fluctuations of hadrons
produced in high-energy heavy-ion collisions can be used as a
measure of whether a quark-gluon system has undergone a phase
transition \cite{bialas}. Until today the question concerning
order of the parton-hadron PT in high-energy collisions is opened.
Lattice gauge calculations indicate that for two flavors the PT
is most likely of the second order \cite{PT_2, PT_2_1}. When
strange quarks are included, it may become a weak first-order PT
\cite{PT_1}.

At present accelerator energies the average number of produced
particles is large enough and therefore the scattering operator
and other operators are inconvenient to consider in terms of number states. At
the same time in such systems as lasers where the average number
of photons is large it is conveniently to use the coherent state
representation (P-representation) \cite{QO_Walls, coher}.
There are a number of efforts to apply P-representation for
investigation of the multiplicity fluctuations as a
phenomenological manifestation of quark-hadron PT in the
framework of the Ginzburg-Landau (GL) formalism both for the
second-order \cite{hwa_nazir, hwa} and first-order \cite{babichev_1, hwa_1,
babichev} PT. In both cases it was supposed that multiplicity distribution
of the hadrons without PT is a Poissonian and scaling behaviour of the factorial
moments has been found. Moreover the scaling exponent $\nu$ is
equal 1.305 for second-order PT \cite{hwa} and $1.32<\nu<1.33$
for the generalized GL model with first-order PT
\cite{babichev}.  There was
the disagreement between experiment and theory. Indeed, NA22 data
on particle production in hadronic collisions give $\nu = 1.45\pm
0.04 $ \cite{Kittel}, heavy-ion experiments $\nu = 1.55\pm 0.12$
\cite{hwa_nazir} and $\nu = 1.459\pm 0.021$ \cite{hevy_ion}.

At the same time the study of multiplicity fluctuations in a
similar type of phase transition have carried out in
nonlinear optics \cite{PT_QO}. It was shown that second-order PT
is related to the symmetry-changing instability of stationary
non-equilibrium states.  This fact may be additional evidence for
the use of optical methods at investigation problems of the PT in the
multiparticle production processes.

Indeed, the idea of applying stochastic methods developed for
studying photon-counting statistics of light beams to particle
production processes was used for explanation of the
experimentally observed properties hadrons, in particular, the
multiplicity distributions, factorial and cumulant moments. It
was shown that the most general distribution that characterizes $
e^+e^-, p\bar p,$ neutrino-induced collisions is a $k$-mode
squeezed state distribution \cite{MD_SS, MD_SS_1}. Squeezed
states (SS) involve  coherent states as the specific case and
posses uncommon properties: they display a specific behaviour of
the factorial and cumulant moments \cite{PSS} and can have both
sub-Poissonian and super-Poissonian statistics corresponding to
antibunching and bunching of photons \cite{QO_Walls},
\cite{QO_Japan}-\cite{Kilin}. Although the SS are constructed in
quantum optics (QO) their relevance to hadron production in
high-energy collisions was recognized long ago \cite{Shih,
Carruthers}. In particular, the  multiplicity distribution of the
pions has been explained by formalism of the squeezed isospin
states \cite{S_isospin_S}. In addition, study of the evolution of
gluon states at the non-perturbative stage of jet development has obtained the new squeezed gluon states \cite{Nasa}-\cite{shaporov} which could be necessary element of hadronization and, in particular, QGP$\rightarrow$hadrons. 
Then using the Local parton hadron
duality it is easy to show that in this case behaviour of hadron
multiplicity distribution in jet events is differentiated from the
negative binomial one. Such specific behaviour of the
multiplicity distribution is confirmed by experiments for $pp,
p\overline{p}$-collisions~\cite{UA5}-\cite{OPAL}.

Considering multiplicity distribution in different collisions
without PT as squeezed one we generalize the coherent state
representation taking into account the squeezed states within
GL-approach for investigation of the multiplicity fluctuations at
phase transition in QGP. Since the multiplicity fluctuations
exhibit intermittency behaviour which is observed in a large
number of experiments, we investigate conditions of appearance of
this effect in depending on the parameters of GL model.

\newpage
\section{Multiplicity distribution of the squeezed states at second-order phase transition}
It is convenient to start description of the
squeezed-state formalism in GL theory with definition of the
photon SS. Two basic kinds of single-mode ideal SS are used in QO:
coherent squeezed state (CSS) and scaling SS (SSS)
\cite{QO_Japan} defined as
\begin{equation}\label{f1}
\left.
\begin{array}{c}
  |\psi, \eta\rangle = D(\psi)S(\eta)|0\rangle \qquad(CSS),\\\\
  |\psi, \eta\rangle = S(\eta)D(\psi)|0\rangle \qquad(SSS),
  \end{array}
  \right\}
\end{equation}
where $D(\psi) = \exp\bigl\{\psi a^+ - \psi^* a\bigr\}$ is a
displacement operator, $\displaystyle S(\eta) =
\exp\Bigl\{\frac{\eta^{*}}{2} a^{2} -
\frac{\eta}{2}(a^{+})^{2}\Bigr\}$ is a squeeze operator, $\psi =
|\psi| e^{i\gamma}$ is an eigenvalue of non-Hermitian annihilation
operator $a$, $|\psi|$ and $\gamma$ are an amplitude and a phase
of the coherent state correspondingly, $\eta = r e^{i\theta}$ is
an arbitrary complex number, $r$ is a squeeze factor, phase
$\theta$ defines the direction of squeezing maximum
\cite{QO_Cambr}. Using general formula for two-photon coherent
state distribution \cite{Yuen} we can write the corresponding
expression for CSS and SSS distributions in the form
\begin{equation}\label{distrib}
 P_n = \frac{1}{\cosh(r)n!}\left(\frac{\tanh(r)}{2}\right)^n \left|H_n(\xi_1)\right|^2
e^{\xi_2},
\end{equation}
where $H_n(\xi_1)$ is a Hermite polynomials, $\xi_1$ and $\xi_2$ are
equal in case of CSS
\begin{equation}\label{css_1}
 \left.
\begin{array}{l}
\xi_1 = \displaystyle
\sqrt{\frac{\langle n\rangle -\sinh^2(r)}{\sinh(2r)}}
\Bigl[\cosh(r) e^{i(\gamma - \theta/2)} + \sinh(r)e^{- i(\gamma - \theta/2)}\Bigr],\\ \\[0.2cm]
\xi_2= \Bigl[\langle n\rangle - \sinh^2(r)\Bigr] 
\Bigl
\{\cosh(2r) [\tanh(r)
\cos(2\gamma - \theta) - 1]
+ \sinh(2r)[\tanh(r) - \cos(2\gamma -
\theta)]\Bigr\}
\end{array}
\right\}
\end{equation}
and for SSS
\begin{equation}\label{sss_1}
 \left.
\begin{array}{l}
\xi_1 = \displaystyle\sqrt{\frac{\langle n\rangle -\sinh^2(r)}{\sinh(2r)}}\;
e^{i(\gamma - \theta/2)}
\Bigl[\cosh(2r) - \sinh(2r) \cos(2\gamma - \theta)\Bigr]^{-\frac12},\\\\
\xi_2 = \displaystyle\frac{[\langle n\rangle -
\sinh^2(r)]\,[\tanh(r)\cos(2\gamma - \theta) -1]}{\cosh(2r) - \sinh(2r)
\cos(2\gamma - \theta)}.
\end{array}
\right\}
\end{equation}
Here $\langle n\rangle$ is an average multiplicity. In particular
case at $\gamma = \theta = 0$ expression (\ref{distrib})
coincides with analogous expressions used for description of the
multiplicity distribution in $ e^+e^-, p\bar p,$ neutrino-induced
collisions \cite{MD_SS, MD_SS_1}.

\begin{figure}[h!]
     \leavevmode
\centering
\includegraphics[width=3.2in, height=2.7in, angle=0]{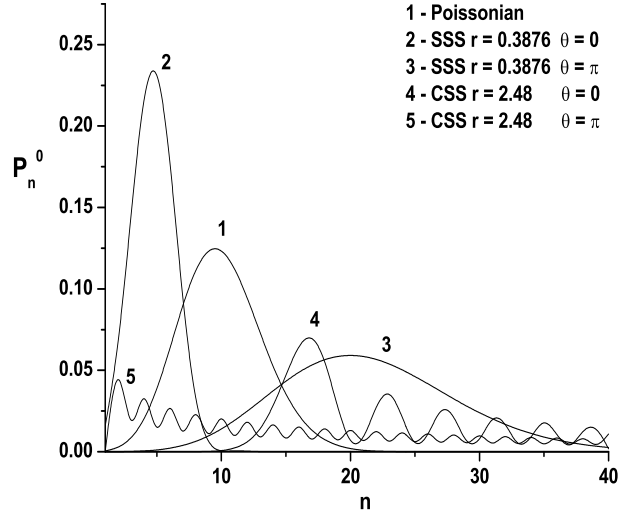}
\caption{Multiplicity distributions before phase transition.}
\label{firstfig}
\end{figure}

In quantum field theory (multi-mode case) the
average number of particles in observed state is defined as \cite{QFT}
\begin{equation}\label{num_1}
  \langle n\rangle = \Bigl\langle \int\limits_V dz a^+(z) a(z) \Bigr\rangle
\end{equation}
and then an average multiplicity for CSS and SSS is equal
correspondingly to
\begin{equation}\label{n}
   \left.
\begin{array}{l}
\langle n\rangle = \displaystyle\int\limits_V|\psi(z)|^2 dz + \sinh^2(r),\\\\
\langle n\rangle = \left(\displaystyle\int\limits_V|\psi(z)|^2 dz \right)\!  \Bigl[\cosh(2r) - \sinh(2r)\cos(2\gamma - \theta)\Bigl] + \sinh^2(r).
\end{array}
\right\}
\end{equation}
Here for simplicity we regard that phase of the coherent state
and of squeezing effect are the same for whole space that is
quantities $\gamma, r, \theta$ are parameters.
\noindent
The probability density of finding $n$ particles in SS is
\begin{equation}\label{num_2}
  |\langle n|\psi(z), \eta\rangle |^2 = P_n^0.
\end{equation}
Then using the expressions (\ref{n}) for average multiplicity we
can write $P_n^0$ in the form
\begin{eqnarray}\label{dens_distr_SS}
  P_n^0 &=&\displaystyle\frac{1}{\cosh(r)n!}\left(\frac{\tanh(r)}{2}\right)^{\!n}
  \displaystyle\left|H_n\left(\left[\int\limits_V|\psi(z)|^2 dz\right]^{\frac12}
  F_1(r,\gamma, \theta)\right)\right|^2\nonumber\\&&{}\times
\exp\biggl\{\displaystyle\int\limits_V|\psi(z)|^2 dz\; F_2(r,\gamma, \theta)\biggr\},
\end{eqnarray}
where $F_1(r,\gamma, \theta), F_2(r,\gamma, \theta)$ are
functions of the parameters $r, \gamma, \theta$ and in
case of CSS are equal to
\begin{equation}\label{F_CSS}
 \left.
\begin{array}{l}
F_1(r,\gamma, \theta) = \displaystyle\frac{\cosh(r) e^{i(\gamma -
\theta/2 )} + \sinh(r) e^{- i(\gamma - \theta/2 )}
}{\sqrt{\sinh(2r)}},\\\\
F_2(r,\gamma, \theta) = \cosh(2r)[\tanh(r) \cos(2\gamma -
\theta) - 1]+ \sinh(2r)[\tanh(r) - \cos(2\gamma - \theta)]
\end{array}
\right\}
\end{equation}
and for SSS
\begin{equation}\label{F_SSS}
 F_1(r,\gamma, \theta) = \displaystyle\frac{e^{i(\gamma - \theta/2
)}} {\sqrt{\sinh(2r)}}\, ,\qquad
F_2(r,\gamma, \theta) = \tanh(r) \cos(2\gamma - \theta) - 1.
\end{equation}
From Fig.1 it is obviously that at $\theta = 0$ we have a sub-Poissonian distribution and at $\theta = \pi$ --- super-Poissonian one. If the squeeze factor is more than one we have oscillations of given distribution (Fig.\ref{firstfig}: CSS distribution).
Obviously that at $r\rightarrow 0$ (the squeezing effect is absent) the
probability density of finding $n$ particles is Poissonian
\begin{equation}\label{Posson}
  P_n^0 = \frac1{n!}\; \exp \biggl\{-\int\limits_V|\psi(z)|^2 dz\biggr\} \left(\int\limits_V|\psi(z)|^2
  dz\right)^{\!\! n}.
\end{equation}

\begin{figure}[h!]
     \leavevmode
\centering
\includegraphics[width=3in, height=2.7in, angle=0]{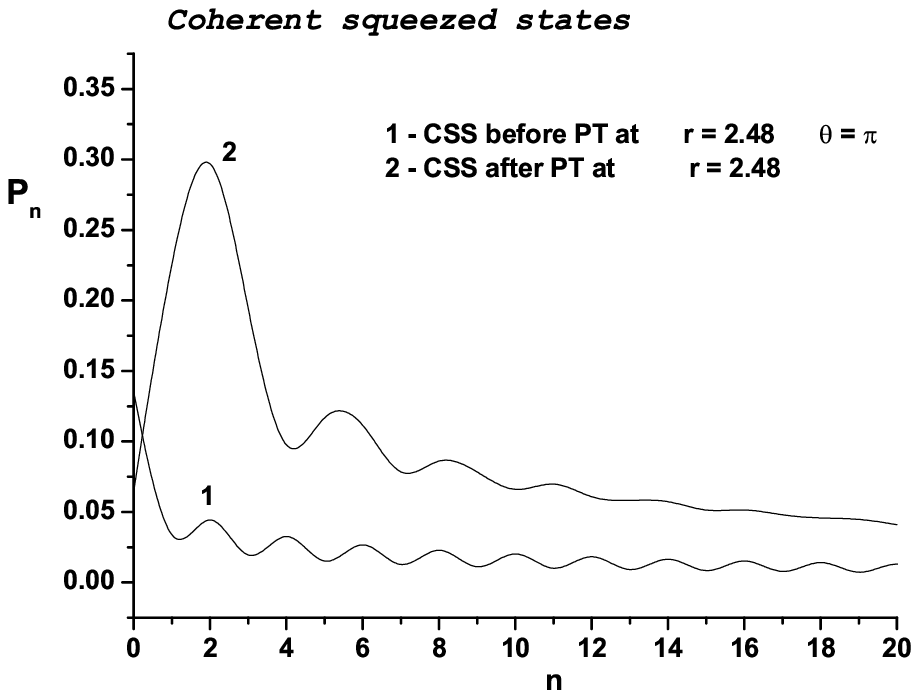}
\includegraphics[width=3in, height=2.7in, angle=0]{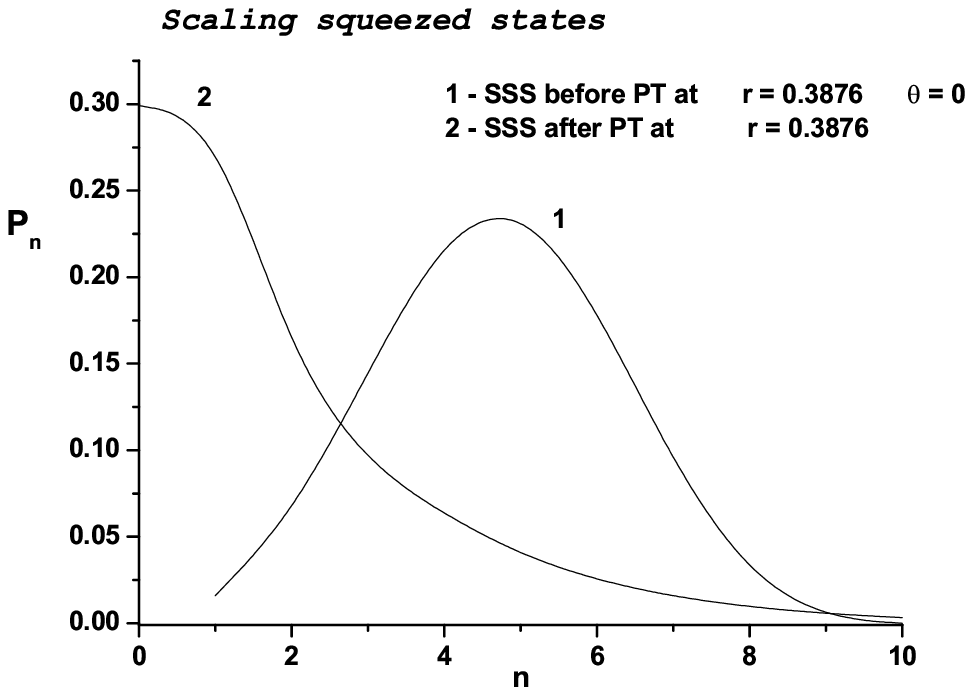}
\caption{Multiplicity distributions of the squeezed states with taking into account of the phase transition.}
\label{secondfig}
\end{figure}

\newpage
Within standard GL model the free energy of the system is
\begin{equation}\label{F}
  F[\psi] = \int dz \{a |\psi(z)|^2 + b|\psi(z)|^4 + c|\partial\psi / \partial z|^2\},
\end{equation}
where $\psi(z)$ is introduced to serve as a complex order parameter.
Then the hadron multiplicity distribution can be given by the
functional integral of the type \cite{stat}
\begin{equation}\label{P_n}
  P_n = Z^{-1} \int D\psi P_n^0 e^{-F[\psi]},
\end{equation}
here $Z = \int D\psi  e^{-F[\psi]}$. Thus the probability of
having a large $n$ in volume $V$ is controled by deviation of
$\psi$ from $\psi_0$ (minimum of the GL potential) as
specified by the thermodynamical factor $e^{-F[\psi]}$. Taking
into account the explicit form of the probability density of
finding $n$ particles in SS $P_n^0$ (\ref{dens_distr_SS}) we
obtain the next expression for the hadron multiplicity
distribution
\begin{eqnarray}\label{P_n_1}
  P_n &=& \frac{1}{2\,Z \cosh(r)}\int D\,\psi\;
  \frac{\tanh^n(r)}{2^n n!}\;
  \exp\left\{- F[\psi]+\int\limits_V|\psi(z)|^2 dz\: F_2(r,\gamma,
  \theta)\right\}\nonumber\\{}&&\times
    \left|H_n\left(\left[\int\limits_V|\psi(z)|^2 dz\right]^{\frac12} F_1(r,\gamma, \theta)\right)
    \right|^2\!\!\!.
\end{eqnarray}
To investigate obtained expression we identify $V=\delta^d$ and regard that $|\psi(z)|$ is constant in every bin width $\delta$ ($d$ is a dimension). Then multiplicity distribution after phase transition is
\begin{eqnarray}\label{P_n_2}
  P_n &=& \frac{1}{2\,\pi \cosh(r)}D^{-1}_{-1}\biggl(-|a|\sqrt{\frac{\delta^d}{2b}}\biggr)
\int\limits_0^{2\pi} d\gamma \; \exp\biggl\{\frac{\delta^d F_2(r,\gamma,
  \theta)(F_2(r,\gamma,\theta)+2|a|)}{8b}\biggr\} \tanh^n(r)\nonumber\\&&{}\times\sum\limits_{k=0}^{n/2}
\sum\limits_{l=0}^{n/2}(-1)^{k+l}\frac{(2k-1)!!\:(2l-1)!!\: n!\:(n-k-l)!}{(2k)!\:(2l)!\:(n-2k)!\:(n-2l)!}\;\left(\frac{2 \delta^d}{b}\right)^{\frac{1}{2}(n-k-l)} \; F_1^{n-2k}(r,\gamma,\theta)\nonumber\\&&{}\times  (F_1^{*})^{n-2l}(r,\gamma,\theta)\, D_{-(n-k-l+1)}\biggl(-\Bigl[|a|+F_2(r,\gamma,\theta)\Bigr]\sqrt{\frac{\delta^d}{2b}}\biggr),
  \end{eqnarray}
where $D_{-f}(w)$ is a function of the parabolic cylinder. Obviously, this expression for $P_n$ is not depended on the phase which defines the direction of squeezing maximum since integrand is a harmonic function of this squeeze parameter. Influence of the phase transition on behaviour of the multiplicity distributions is shown on Fig\footnote{Values of the parameters $a, b, r$ correspond to case then we have intermittency and the scaling exponent value is equal to 1.459.}.\ref{secondfig}.

This expressions for $P_n$ (\ref{P_n_1}), (\ref{P_n_2}) will be essential at analysis phenomenon of intermittency.

\section{Intermittency}
One of the effective way to manifest the nature of the
multiplicity fluctuations in high-energy collisions is to examine
the dependence of the normalized factorial moments $F_q$ \cite{Kittel, intermit}
\begin{equation}\label{F_q}
  F_q = \frac{\langle n(n-1)\cdot\cdot\cdot (n-q+1)\rangle}{\langle n\rangle ^q} =
  \frac{f_q}{f_1^q}
\end{equation}
on the bin width $\delta$ in rapidity. Here $f_q = \langle n~(n~-~1~)\cdot~\cdot~\cdot~ (n-q+1)~\rangle,$ $n$
is the number of hadrons detected in $\delta$ in an event, and
the average are taken over all events. The multiplicity
fluctuations can exhibit intermittency behaviour which is
manifested by power-law behaviour of $F_q$ on $\delta$
\cite{Kittel}
\begin{equation}\label{intermit_1}
  F_q\propto \delta^{-\varphi_q},
\end{equation}
where $\varphi_q$ is referred to as the intermittency index.
Indeed, apart from collision energy and nuclear size we can vary
only the size of a cell $\delta$ in phase space that is just the
central theme of intermittency. This effect has been observed in
a large number of experiments: $e^+e^-, \mu p, pp, pA$ and $AA$
collisions \cite{Kittel}.

\begin{figure}[h!]
     \leavevmode
\centering
\includegraphics[width=2.7in, height=2.3in, angle=0]{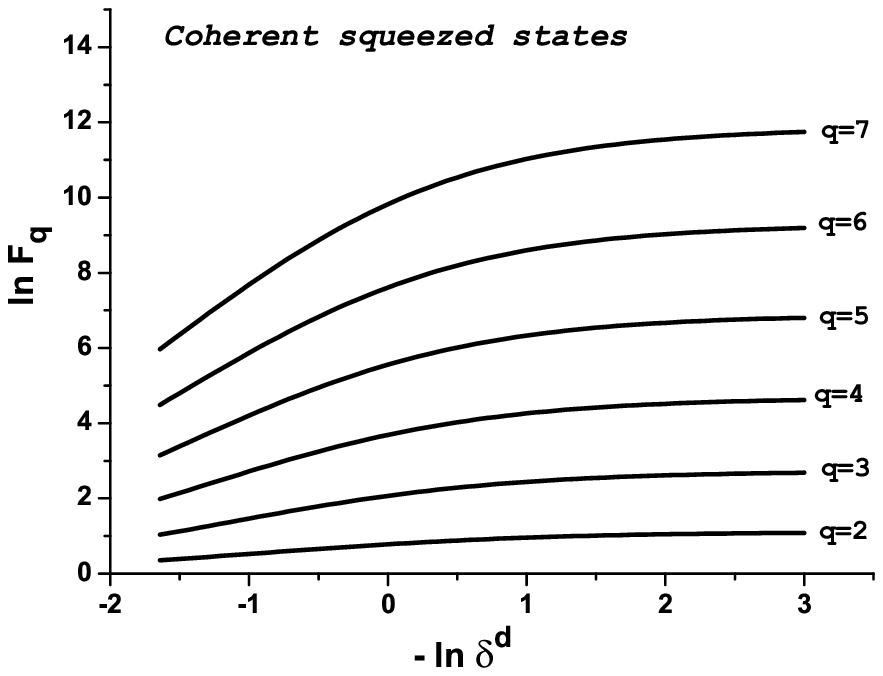}
\includegraphics[width=2.7in, height=2.3in, angle=0]{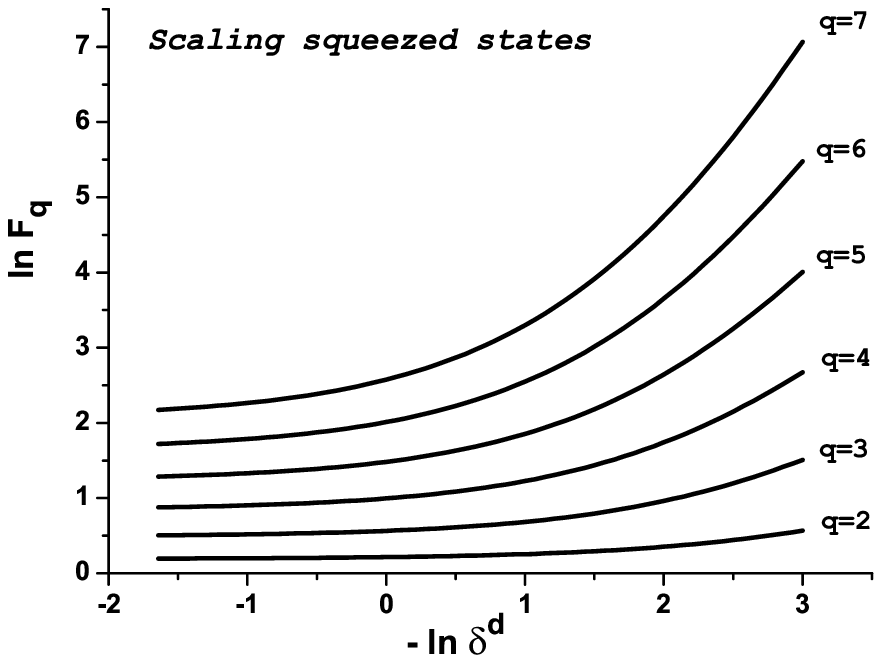}
\caption{Log-log plot of $F_q$ vs $1/\delta^d$ in case of the squeezed states.}
\label{thirdfig}
\end{figure}

\begin{figure}[h!]
     \leavevmode
\centering
\includegraphics[width=2.7in, height=2.3in, angle=0]{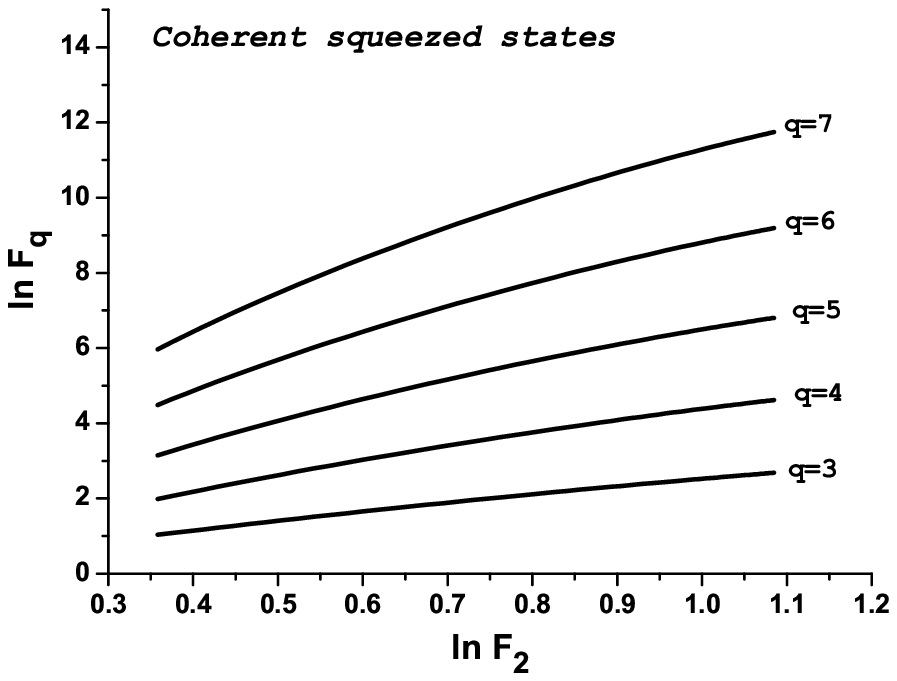}
\includegraphics[width=2.7in, height=2.3in, angle=0]{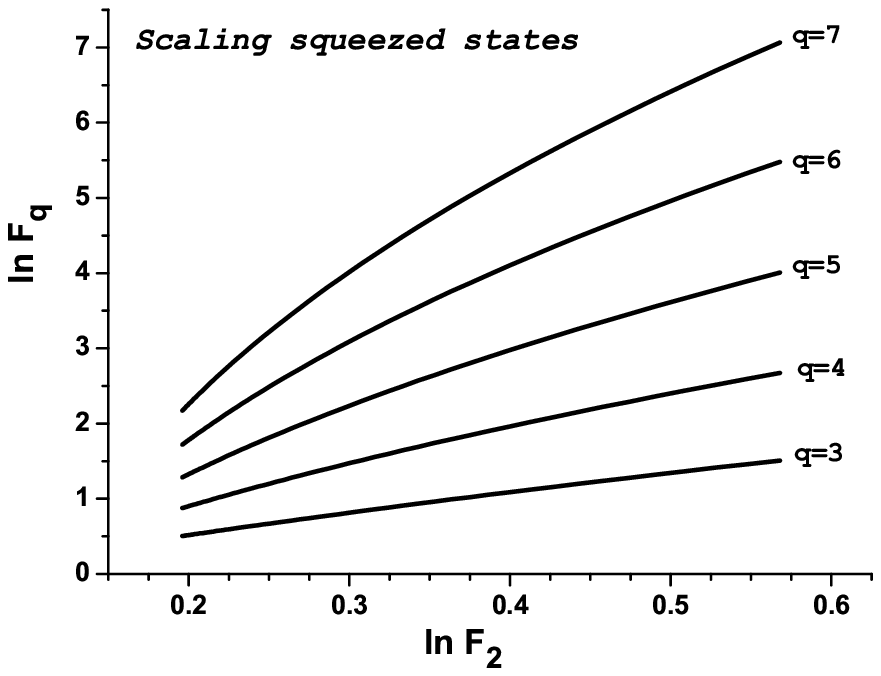}
\caption{Log-log plot of $F_q$ vs $F_2$ in case of the squeezed states.}
\label{forthfig}
\end{figure}

Therefore the intermittency analysis is used to explore universal
characteristics of quark-hadron PT in the GL model. In this
section we examine whether (\ref{intermit_1}) is valid under taking
into account PT. Since
\begin{equation}\label{f_q}
  f_q =  \sum\limits_{n=q}^{\infty}\frac{n!}{(n-q)!}P_n,
\end{equation}
using (\ref{P_n_1}) and (\ref{F}) we obtain the next explicit
form of $f_q$
\begin{equation} \label{f_q_1}
 \begin{array}{l}
f_q = \displaystyle\frac1{2\,Z \cosh(r)}
\int D\psi\;
e^{- F[\psi]}\exp\left\{\displaystyle\int
\limits_V|\psi(z)|^2 dz\: F_2(r,\gamma,
\theta)\right\}\\\vspace{-0.3cm}\\\hspace{0.7cm}\times\displaystyle
\sum\limits_{n=q}^{\infty}
\frac{1}{(n-q)!}
\left(\frac{\tanh(r)}{2}\right)^{\!n}
\displaystyle
\left|H_n\left(\left[\int\limits_V|\psi(z)|^2 dz\right]^{\frac12} F_1(r,\gamma,
  \theta)\right)\right|^2,
\end{array}
\end{equation}
where $V$ is the volume of the cell in which the factorial moment
is measured. Taking into account the formula \cite{ryad}
\begin{equation}\label{hhhh}
 \begin{array}{l}
  \displaystyle\sum\limits_{k=0}^{\infty}\displaystyle\frac{t^k}{k!} H_{k+m}(x)
  H_{k+n}(y)=
  (1-4t^2)^{-(m+n+1)/2}\exp\left[\displaystyle
  \frac{4xyt - 4 t^2(x^2+ y^2)}{1-4t^2}\right]
  \\\vspace{-0.3cm}\\\vspace{-0.3cm}\\\qquad{}\times
  \displaystyle\sum\limits_{k=0}^{min(m,n)} 2^{2k} k! \left( {{\begin{array}{*{20}c}
 m \hfill \\
 k \hfill \\
\end{array} }} \right)\left( {{\begin{array}{*{20}c}
 n \hfill \\
 k \hfill \\
\end{array} }} \right) t^k
  H_{m-k}\left(\displaystyle\frac{x-2ty}{\sqrt{1-4t^2}}\right) H_{n-k}
  \left(\displaystyle\frac{y-2tx}{\sqrt{1-4t^2}}\right)
\end{array}
\end{equation}
and identifying $V=\delta^d$, regarding that $|\psi(z)|$ is
constant in every bin width $\delta$, we rewrite the expression
(\ref{f_q_1}) taking into account the explicit form of
$F_1(r,\gamma, \theta), F_2(r,\gamma, \theta)$  for CSS
(\ref{F_CSS}) and SSS (\ref{F_SSS}) correspondingly in the next form
\begin{equation}\label{f_q_21}
 \begin{array}{l}
 {\rm (CSS)}\\
   f_q = \displaystyle(2\,Z)^{-1}\sinh^{2q}(r)\int\limits_0^{2 \pi} d\gamma
  \int\limits_0^{\infty}d |\psi|^2 e^{-F[\psi]}
\displaystyle
\sum\limits_{n=0}^{q}\left(\displaystyle\frac{q!}{n!}
\right)^2\displaystyle\frac1{(q-n)!
(2\tanh(r))^n}\\\vspace{-0.3cm}\\\qquad{}\times\left|H_n\left(\sqrt{
\displaystyle\frac{|\psi|^2 \delta^d}{\sinh(2r)}}\; e^{i(\gamma -\theta/2 )}
\right)\right|^2,
\end{array}
\end{equation}
\begin{equation}\label{f_q_2}
 \begin{array}{l}
 {\rm (SSS)}\\
   f_q = \displaystyle(2\,Z)^{-1}\sinh^{2q}(r)\int\limits_0^{2 \pi} d\gamma
  \int\limits_0^{\infty}d |\psi|^2 e^{-F[\psi]}
\displaystyle\sum\limits_{n=0}^{q}\left(\displaystyle\frac{q!}{n!}\right)^2
\displaystyle\frac1{(q-n)!
  (2 \tanh(r))^n}\\\vspace{-0.3cm}\\\qquad{}
\times\Biggl|H_n\Biggl(
\sqrt{\displaystyle\frac{|\psi|^2 \delta^d}
  {\sinh(2r)}} \left[\cosh(r) e^{i(\gamma -\frac{\theta}2)} - \sinh(r)
  e^{- i(\gamma - \frac{\theta}2)}
  \right]
  \Biggr)
  \Biggr|^2.
\end{array}
\end{equation}
Integrating obtained expressions we can represent their as
\begin{equation}\label{f_q_3}
  f_q = \frac{J_q}{J_0},
\end{equation}
where in case CSS
\begin{eqnarray}\label{J_q_CSS}
 J_q &=& \frac{\pi}{\sqrt{2b\delta^d}}\exp\biggl\{\frac{|a|^2 \delta^d}{8b}\biggr\}\sinh^{2q}(r)\sum\limits_{n=0}^q \frac{(q!)^2}{(q-n)!}\tanh^{-n}(r)\sum\limits_{k=0}^{n/2}
\biggl(\frac{(2k-1)!!}{(2k)!}\biggr)^{\! 2} \nonumber\\&&{}\times\frac{(\sinh(2r))^{2k-n}}{(n-2k)!}\left(\frac{2 \delta^d}{b}\right)^{\!\frac{1}{2}(n-2k)} \; D_{-(n-2k+1)}\biggl(-|a|
\sqrt{\frac{\delta^d}{2b}}\biggr)
\end{eqnarray}
and for SSS
\begin{eqnarray}\label{J_q_SSS}
 J_q &=& \frac{\pi}{\sqrt{2b\delta^d}}\exp\biggl\{\frac{|a|^2 \delta^d}{8b}\biggr\}\sinh^{2q}(r)\sum\limits_{n=0}^q \frac{(q!)^2}{(q-n)!}\tanh^{-n}(r)\sum\limits_{k=0}^{n/2}\sum\limits_{l=0}^{n/2}
\frac{(2k-1)!!\;(2l-1)!!}{(2k)!\;(2l)!} \nonumber\\&&{}\times (\sinh(2r))^{k+l-n}\; (n-k-l)! 
\left(\frac{2 \delta^d}{b}\right)^{\!\frac{1}{2}(n-k-l)}\;
D_{-(n-k-l+1)}\biggl(-|a|
\sqrt{\frac{\delta^d}{2b}}\biggr)\nonumber\\&&{}\times\sum\limits_{j=0}^{n-2l}
\frac{(\sinh(r))^{l-k+2j}\; (\cosh(r))^{2n-k-2l-2j}}{j!\;(l-k+j)!\;(n-k-l-j)!\;(n-2l-j)!}.
\end{eqnarray}
Then according to (\ref{F_q}),(\ref{f_q_3}) the normalized factorial moments $F_q$
have the next form
\begin{equation}\label{F_q_1}
  F_q = J_q J_1^{-q} J_0^{q-1}.
\end{equation}
On Fig.\ref{thirdfig} and Fig.\ref{forthfig} we represent the results of analysis of the dependences of $\ln F_q$ on $(-\ln\delta^d)$ and of $\ln F_q$ on $(\ln F_2)$
correspondingly for the squeeze factors $r = 2.48$ (CSS) and $r = 0.3876$ (SSS) and for the
next values of the parameters of the GL model $a = -10, b =
0.20055$.


\begin{figure}[h!]
     \leavevmode
\centering
\includegraphics[width=3in, height=2.3in, angle=0]{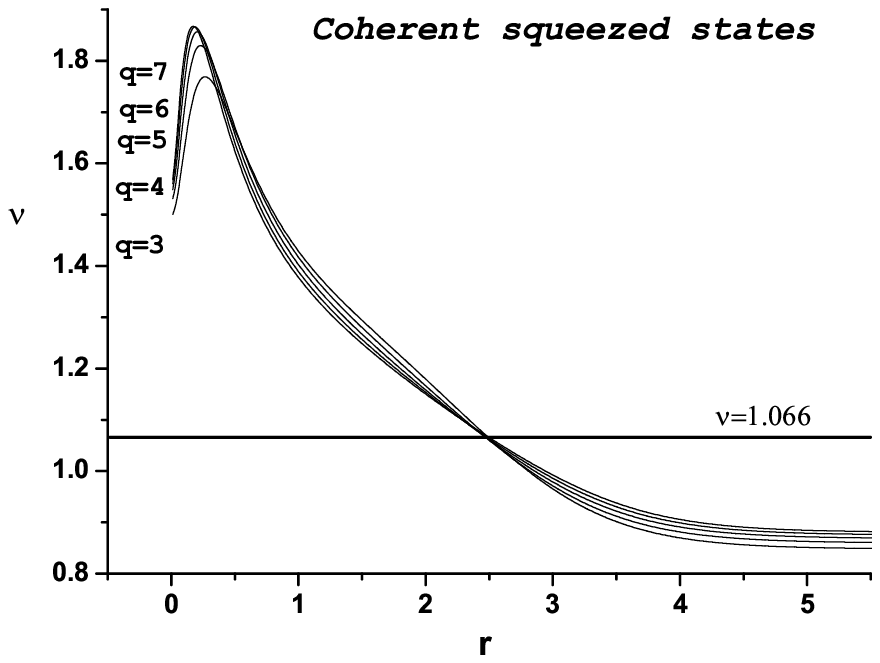}
\includegraphics[width=3in, height=2.3 in, angle=0]{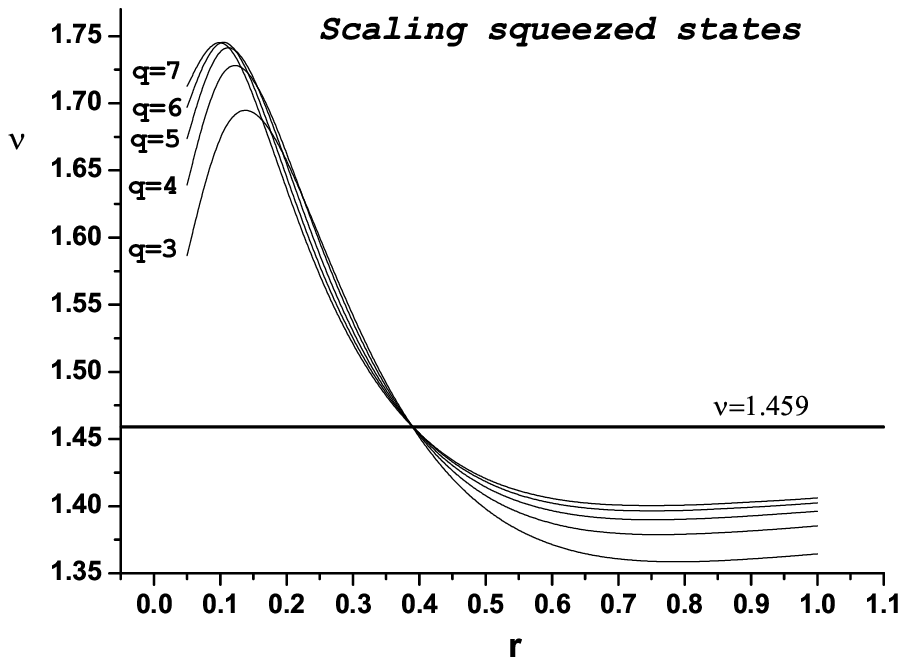}
\caption{Dependence of the scaling exponent on squeeze factor $r$.}
\label{fifthfig}
\end{figure}

If the local slope of $\ln F_q$ vs $\ln F_2$ is approximately
constant then we would have the scaling behaviour (Ochs-Wosiek
scaling law) \cite{Ochs-Wosiek}
\begin{equation}
F_q \propto {F_2}^{\beta_q},
\end{equation}
which is valid for intermittent systems \cite{intermit}. The
slopes $\beta_q$ are well fitted by the formula \cite{hwa_nazir}
\begin{equation}
\beta_q = (q-1)^\nu,
\end{equation}
where $\nu$ is a scaling exponent. Dependence of $\nu$ on squeeze
factor $r$  is represented on the Fig.\ref{fifthfig} at the same values of the
parameters of the GL model.

It is obvious from Fig.\ref{fifthfig} that we have intermittency in case of the CSS when $\nu=1.066$ at $r=2.48$ and $\nu=1.459$ at $r=0.3876$ for SSS. Thus scaling behaviours of the
normalized factorial moments for the scaling squeezed states are characterized by obtained
scaling exponent $\nu$ that agrees with experimental NA22 data on
particle production in heavy-ion experiments \cite{hevy_ion}. Parameters at which the scaling exponent values agree with various experimental data \cite{Kittel, hwa_nazir} are represented in the Tab.\ref{tab1}.
\begin{table}[htbp]
\begin{center}
\caption{Parameters at which the scaling exponent values agree with experimental data.}

\par
\begin{tabular}
{|p{20pt}|p{45pt}|p{45pt}|p{45pt}|p{45pt}|p{45pt}|p{45pt}|}
\hline
\raisebox{-1.50ex}[0cm][0cm]{\hspace{0.1cm} a}& 
\multicolumn{2}{|p{91pt}|}{$\quad\;\;\nu = 1.450$} & 
\multicolumn{2}{|p{91pt}|}{$\quad\;\;\nu = 1.459$} & 
\multicolumn{2}{|p{91pt}|}{$\quad\;\;\nu = 1.550$}  \\
\cline{2-7} 
 & 
\centering b& 
\centering r& 
\centering b& 
\centering r& 
\centering b& 
\hspace{0.7cm} r \\
\hline
-1& 
0.00663& 
0.32363& 
0.00592& 
0.32663& 
0.00220& 
0.34375\\
\hline
-2& 
0.02761& 
0.31312& 
0.02455& 
0.31826& 
0.00992& 
0.32266 \\
\hline
-3& 
0.05217& 
0.32917& 
0.04988& 
0.32216& 
0.02275& 
0.31387 \\
\hline
-4& 
0.07313& 
0.35251& 
0.06956& 
0.34795& 
0.03760& 
0.31758 \\
\hline
-5& 
0.09535& 
0.36548& 
0.09052& 
0.36190& 
0.05582& 
0.31572 \\
\hline
-6& 
0.11815& 
0.37385& 
0.11213& 
0.37079& 
0.07140& 
0.32305 \\
\hline
-7& 
0.14138& 
0.37970& 
0.13398& 
0.37693& 
0.08570& 
0.33040 \\
\hline
-8& 
0.16472& 
0.38401& 
0.15609& 
0.38144& 
0.09871& 
0.33740 \\
\hline
-9& 
0.18819& 
0.38730& 
0.17834& 
0.38489& 
0.11190& 
0.34261 \\
\hline
-10& 
0.21187& 
0.38983& 
0.20055& 
0.38758& 
0.12518& 
0.34664 \\
\hline
\end{tabular}
\label{tab1}
\end{center}
\end{table}

In case of the CSS  the scaling exponent values are not agree with various experimental data \cite{Kittel, hwa_nazir} at any values of the parameters $a, b, r$.

\section{Conclusion}
We study multiplicity fluctuations and intermittency in second order phase transition from QGP to hadrons within of the GL model. Generalizing P-representation to squeezed state one (in particular, for two types: CSS, SSS) we obtain the explicit expressions for the probability of finding $n$ particles and for the normalized factorial moments $F_q$ which include additional parameters $r, \theta$ inherent to the squeezing effect.

Changing new parameters we can more successfully apply GL model for description of the phase transitions. Indeed, at $a = -10, b = 0.20055$ and at $r=0.3876$  in case of the scaling squeezed states of the hadrons we have intermittency when the value of the scaling exponent is equal to 1.459. Obtained value of the scaling exponent
agrees with experimental data \cite{hwa_nazir, Kittel, hevy_ion}.

We hope that squeezed state approach will be available for description of fluctuations in the phase transition from quark-gluon plasma to hadrons in processes where
an energy density is very high, for example, in heavy ion collisions
at high energy.

\end{document}